\documentclass[twocolumn,showpacs,prl]{revtex4}
\usepackage{amssymb}

\usepackage{graphicx}

\begin{document} 
\title{Slow light in degenerate Fermi gases}
\author{G. Juzeli\={u}nas$^1$ and P. \"Ohberg$^2$}

\affiliation{$^1$Vilnius University Research Institute of Theoretical Physics and
Astronomy,\\
A. Go\v{s}tauto 12, 2600 Vilnius, Lithuania\\
$^2$Department of Physics, University of Strathclyde,\\
Glasgow G4 0NG, Scotland}

\begin{abstract}
We investigate the effect of slow light propagating in a degenerate atomic Fermi gas. 
In particular we use slow light with an orbital angular momentum. We
present a microscopic theory for the interplay between light and matter and
show how the slow light can provide an effective magnetic field acting on the 
electrically neutral fermions, a direct
analogy of the free electron gas in an uniform magnetic field. As an example
we illustrate how the corresponding de Haas-van Alphen effect can be seen in
a gas of neutral atomic fermions.
\end{abstract}


\maketitle


The recent advances in trapping and cooling atoms has provided an excellent
starting point for studying many different types of physical phenomena,
ranging from fundamental atomic physics to cosmological aspects \cite{sta03}. 
In this respect atomic Bose-Einstein condensates have attracted a lot of interest 
\cite{bec_stri}. Recently several experimental groups have succeeded in
trapping and cooling also fermions \cite{demarco99,ketterle03} well below
the Fermi temperature. Fermi systems are well known from the study of
electron properties in materials. Trapped atomic fermions are electrically neutral and a
direct analogy between the magnetic properties of these systems and solid
state phenomena is not necessarily straightforward. We suggest this problem
can be circumvented if the properties of slow light is used, i.e. light with
a group velocity as low as meters per second \cite{hau,kash99,budker99}. The coupling
between the slow light and the atoms can give rise to some remarkable
effects such as dragging of the light \cite{ulf+paul98,artoni03,juz+mas+fle03} and
complete coherent freezing of the pulse \cite{fleisch+lukin00,freeze,freeze1}. 
In a similar manner the slow light should affect the atomic motion.


In this Letter we investigate the influence of slow light on the mechanical properties 
of a degenerate Fermi gas of atoms. The theory is 
fully microscopic and based on the explicit analysis of the quantum 
dynamics of atomic fermions
coupled to the electromagnetic field. In particular we use slow light with
an orbital angular momentum \cite{allen99,oam}. This allows us to introduce
an effective magnetic field which acts on the electrically neutral fermions. As such we
have a typical, often regarded as an academic, text book scenario with
free electrons moving in a constant magnetic field. This opens up the
possibility to study phenomena well known from solid state and condensed
matter physics, with all the benefits given by the trapped atoms where a
range of experimental parameters such as atom-atom interactions, particle
numbers, the shape of the trapping potential etc. can easily be manipulated.
In addition, using light as the effective magnetic field is going to be
favourable since it is rather difficult to control real magnetic fields. As an example we show how 
the de Haas-van Alphen effect is obtained in a
neutral cloud of fermions. Finally we conclude by discussing some other
aspects and possibilities with slow light in degenerate quantum gases. 


\begin{figure}[tbp]
\center{
\includegraphics[width=7cm]{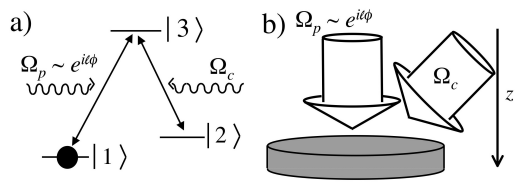}
\caption{a) The level scheme for the Electromagnetically  Induced Transparency with the probe beam $\Omega_p$and 
control beam $\Omega_c$. b) Schematic representation of the experimental setup with the two light beams 
incident on the cloud of atoms. The probe beam propagates in the $z$-direction. The control beam can propagateparallel 
\cite{freeze,freeze1}, perpendicular \cite{hau} or antiparallel to the probe beam. }}
\end{figure}

Light can be slowed down \cite{hau,kash99,budker99} by using the properties of
Electromagnetically Induced Transparency (EIT) \cite{Arimondo96,Harris97,eit,lukin03}, 
in which the group velocity of the light, $v_g$, is reduced by applying another beam, called the control beam. The beams act on the $\Lambda$-type atoms characterized by two
hyper-fine ground levels $1$, and $2$, as well as an electronic excited
level $3$, as depicted in Fig. (1). Initially the atoms occupy the lowest level 
$1$. 

The atoms are described in terms of the fermionic field-operators $\Psi _{j}\left(\mathbf{r},t\right)$ 
representing the second-quantized wavefunction for the translational motion of atoms in the 
$j$-th electronic state, with $j=1,2,3$. The operator $\Psi _{j}\left(\mathbf{r},t\right)$  
annihilates an atom positioned at $\mathbf{r}$ characterized by the internal state $j$. 
In what follows, the spatial and 
temporal variables will be kept implicit in
$\Psi _{j} \left( \mathbf{r},t\right) \equiv \Psi _{j} $. 

The atoms interact with two laser beams: A strong control laser drives the
transition $\left| 2\right\rangle \longrightarrow \left| 3\right\rangle $,
whereas a weaker probe field is associated with the transition 
$\left| 1\right\rangle \longrightarrow \left| 3\right\rangle $ (see Fig.(1)). The
control laser has a frequency $\omega _{c}$, a wave-vector $\mathbf{k}_{c}$,
and a Rabi frequency $\Omega_c =\Omega _{c}^{(0)}\exp \left( i\mathbf{k}_{c}.
\mathbf{r}\right)$, where $\Omega_c^{(0)}$ is a slowly varying amplitude. The 
probe field, on the other hand, is characterized by a central frequency 
$\omega_p =ck_p$, a wave-vector $\mathbf{k}_p=k_p\mathbf{\hat{z}}$, and a Rabi
frequency 
\begin{equation}
\Omega_p=\Omega_p^{(0)} e^{i(\ell \phi+\mathbf{k}_{p}\cdot{\mathbf r})} \label{omega_p}
\end{equation}
where $\Omega _{p}^{(0)}$ is a slowly varying amplitude. In writing Eq.(\ref
{omega_p}) we have allowed the probe photons to have an orbital angular
momentum $\hbar \ell$ along the \textbf{z} axis \cite{allen99,oam}.

Introducing the slowly-varying atomic field-operators $\Phi _{1}=\Psi
_{1}e^{i\omega _{1}t}$, $\Phi _{3}=\Psi _{3}e^{i\left( \omega _{1}+\omega_p
\right) t}$ and $\Phi _{2}=\Psi _{2}e^{i\left( \omega _{1}+\omega_p -\omega
_{c}\right) t}$, and adopting the rotating wave
approximation, one can write the following equations of motion:
\begin{eqnarray}
\!\!\!\!\!i\hbar \dot{\Phi}_{1}&=&-\frac{\hbar ^{2}}{2m}\nabla ^{2}\Phi
_{1}+V_{1}( \mathbf{r})\Phi _{1}+\hbar \Omega_p^*\Phi _{3},  \label{eq-at-g2}
\\
\!\!\!\!\!i\hbar \dot{\Phi}_{3}&=&\!\!\!(\epsilon _{31}\!-\!\frac{\hbar ^{2}%
}{2m}\nabla ^{2}) \Phi _{3}\!+\!V_{3}(\mathbf{r})\Phi _{3}\!+\!\hbar
\Omega_c \Phi _{2}\!+\!\hbar \Omega_p\Phi _{1}  \label{eq-at-e2} \\
\!\!\!\!\!i\hbar \dot{\Phi}_{2}&=&\!\!\!( \epsilon _{21}\!-\!\frac{\hbar ^{2}}{2m%
}\nabla ^{2}) \Phi _{2}\!+\!V_{2}(\mathbf{r})\Phi _{2}\!+\!\hbar \Omega_c ^{\ast
}\Phi _{3},  \label{eq-at-q2}
\end{eqnarray}
where $m$ is the atomic mass, $V_{j}(\mathbf{r})$ is the trapping potential for an atom in
the electronic state $j$, 
$\epsilon _{21}=\hbar \left( \omega _{2}-\omega_{1}+\omega _{c}-\omega_{p} \right) $ 
and 
$\epsilon _{31}=\hbar \left( \omega
_{3}-\omega _{1}-\omega _{p} \right) $ are, respectively, the energies of the
detuning from the two- and single-photon resonances, 
$\hbar \omega_{j}$ being the electronic energy of the atomic level $j$. 

It is noteworthy that the dissipation of the excited electronic state can be included into
equation (\ref{eq-at-e2}) replacing $\epsilon _{31}$ by $\epsilon
_{31}-i\hbar \gamma _{31}$ and adding the appropriate noise operator.
The second hyperfine state $\left| 2\right\rangle$ has usually 
a small decay rate $\gamma _{21}$ which can therefore be omitted in the 
corresponding equation (\ref{eq-at-q2}).  Note also that  
the equations of motion (\ref{eq-at-g2})-(\ref{eq-at-q2}) do not
accommodate collisions between the ground-state atoms. This is legitimate
for the degenerate Fermi gas in which s-wave scattering is forbidden and
only weak p-wave scattering is present \cite{butts97,demarco99,mewes00,juzeliunas01}.

Since the probe field is much weaker than the control field ($|\Omega_p| \ll
|\Omega_c| $), depletion of the ground-state atoms is small. Furthermore, we
assume that the two-photon detuning $\epsilon _{21}$ is sufficiently
small. Neglecting the terms with $\Phi _{3}$, $\nabla ^{2}\Phi _{3}$\ and $%
\dot{\Phi}_{3}$ in Eq. (\ref{eq-at-e2}), one arrives at the adiabatic
condition \cite{Arimondo96,Harris97,eit,lukin03} 
relating $\Phi _{2}$ to $\Phi _{1}$ as: 
\begin{equation}
\Phi _{2}(\mathbf{r},t)=-\zeta \Phi _{1}(\mathbf{r},t).  \label{psi-q1}
\end{equation}
where $ \zeta \equiv \zeta({\bf r}) = \Omega_p/\Omega_c $.
The condition (\ref{psi-q1}) implies that the control and probe beams have driven the 
atoms to the dark state 
$\left| 1\right\rangle - \zeta \left| 2\right\rangle $ representing a special superposition between the two hyperfine 
ground states \cite{Arimondo96,Harris97,eit,lukin03}.
If the atoms are in the dark state, the resonant control and probe beams can not populate the upper atomic level $3$,
as the two beams contribute destructively to the absorption process.
This justifies neglecting the decay of the upper atomic level $3$
in the equation of motion (\ref{eq-at-e2}). 

Equation (\ref{psi-q1}) shows that the orbital angular momentum $\hbar \ell$ of the probe field 
$\Omega_p \thicksim e^{i\ell\phi}$
is transferred into the orbital angular momentum of the centre of mass motion for 
atoms occupying level $2$.
This goes along with a general rule saying that the 
exchange of the orbital angular momentum in the electric dipole approximation occurs 
exclusively between the light and the atomic centre of mass motion \cite{mohamed02}. 
The rule has been implicitly assumed in the initial equations of motion 
(\ref{eq-at-g2})-(\ref{eq-at-q2}) containing no contributions due to exchange in the 
orbital angular momentum between the internal atomic states and the centre of mass motion.    


Consider now the influence of the slow light on the dynamics of the ground state atoms.
Using Eqs.(\ref{eq-at-q2}) and (\ref{psi-q1}), one has: 
\begin{equation}
\Phi _{3}(\mathbf{r},t)=-\frac{1}{\hbar \Omega_c ^{\ast }}\left( \frac{\hbar
^{2}}{2m}\nabla ^{2}+i\hbar \frac{\partial }{\partial t}-\epsilon
_{21}-V_{2}(\mathbf{r})\right) \left( \zeta \Phi _{1} \right) .
\label{psi-e1}
\end{equation}
The relationships (\ref{eq-at-g2})\ and\ (\ref{psi-e1}) provide the
following equation of motion for the field operator $\Phi _{1}({\bf r})$, 
\begin{equation}
i\hbar \dot{\Phi}_{1}=\frac{1}{2m}\left[ i\hbar \mathbf{\nabla }+\mathbf{A}
_{eff}\right] ^{2}\Phi _{1}+V_{eff}(\mathbf{r})\Phi _{1},  \label{Eq-psi-1}
\end{equation}
where 
\begin{equation}
{\bf{A}}_{eff} ({\bf r})=i\hbar \zeta ^{\ast } \mathbf{\nabla } \zeta \equiv 
- \hbar |\zeta|^2 \mathbf{\nabla } S  +
i\frac{\hbar}{2} \mathbf{\nabla }|\zeta |^2 
\label{A-eff-rez}
\end{equation}
and 
\begin{equation}
V_{eff}(\mathbf{r})=V_1 (\mathbf{r}) 
+ \left( \left| \zeta \right| ^{-2} - 2 \right)  \frac{ \left| \mathbf{A}_{eff}\right| ^{2}}{2m}
+\hbar \omega _{21}\left| \zeta \right| ^{2}  \label{V-eff-rez}
\end{equation} 
are the \textbf{effective vector} and \textbf{trapping potentials}, with 
$\hbar \omega _{21}=\epsilon _{21}+V_{2}(\mathbf{r})-V_{1}(\mathbf{r})$. 
Here the dimensionless function 
$\zeta = e^{iS} \Omega_p^{(0))}/\Omega_c^{(0)}$
is characterized by a phase $S=(\mathbf{k_p}-\mathbf{k_c})\cdot\mathbf{r}+ \ell\phi $.   
In writing Eqs. (\ref{Eq-psi-1})-(\ref{V-eff-rez}), 
we made use of the assumprion $\left| \zeta \right|^2 \ll 1$. 
Note that such an assumption is not essential in 
deriving Eqs. (\ref{Eq-psi-1})-(\ref{V-eff-rez}).
By relaxing the condition $\left| \zeta \right|^2 \ll 1$, one arrives at
${\bf{A}}_{eff} =i\hbar \left( 1+\left| \zeta \right| ^{2}\right) ^{-1}
\zeta ^{\ast } \mathbf{\nabla } \zeta $.
In such a situation, $V_{eff}$ also experiences modifications. This effect will be explored elsewhere.


It is instructive to note that 
$\mathbf{A}_{eff}$ is generally non-Hermitian. The
Hermitian contribution is due to the changes in the phase of $\zeta$, the
non-Hermitian one being due to the changes in the amplitude. The non-Hermitian
part of $\mathbf{A}_{eff}$ can be eliminated by a pseudo gauge
transformation $\Phi _{1} = \Phi _{1}^{(0)} \exp[-|\zeta|^2/2]$, where the
transformed operator $\Phi _{1}^{(0)}$ undergoes a unitary evolution. 
Since $|\zeta|^2\ll1$, one can neglect the small changes in the
amplitude of $\Phi _{1}$ making the operator $\mathbf{A}_{eff}$ Hermitian.
Note also that the probe field $\Omega_p$ is considered to be
an incident quantity not affected by the induced motion of the ground-state fermions. 
Consequently the probe field $\Omega _{p}$ 
undergoes
a usual propagation at 
a group velocity $v_{g}\sim \left| \Omega _{c}\right| ^{2}$ 
\cite{Arimondo96,Harris97,eit,lukin03} in the $z$-direction.


In this way, we can create an effective vector potential through the
phase $S$ of the incoming probe beam. The experimental situation is 
schematically described in Fig. (1) where the incoming probe
beam is of the form $e^{i \ell \phi}$. Suppose that the intensity of the control field does not
vary considerably within the atomic cloud. If we consider co-propagating control
and probe beams and choose the intensity of the probe beam of the form 
$|\Omega_p|^2 \sim r^2$ in the transversal plane, we obtain the effective
vector potential 
\begin{equation}
\mathbf{A}_{eff} \equiv - \hbar \ell |\zeta|^2 \mathbf{\nabla } \phi  
=\frac{\hbar \ell \alpha_0}{R^2} (y \hat e_x-x \hat e_y)
\label{eff}
\end{equation}
where $\alpha_0=|\zeta|^2 R^2/r^2$ is a ratio (typically $\alpha_0 < 0.1$)
between the intensities of the probe and the control beam at radius $R$ 
of a cylinder in which the gas is contained. Such an external trap can be created by for
instance high order Bessel beams \cite{arlt01,wright01}. It is interesting
to note here that with this choice of light the effective vector potential (\ref{eff})
corresponds to a constant magnetic field in the direction opposite to the z-axis, since we have
the relation 
\begin{equation}
\mathbf{B}_{eff}=\nabla\times \mathbf{A}_{eff}= -\frac{2\hbar \ell \alpha_0}{R^2} \hat e_z.
\end{equation}
The strength of the effective magnetic field is given by the orbital angular
momentum of the light and can be controlled by applying suitable phase and
intensity holograms \cite{oam}. It is relatively straightforward to create
and control high angular momenta of the order of several hundred $\ell$ which
consequently controls the effective magnetic field. 

Equation (\ref{Eq-psi-1}) describes the effective dynamics of trapped noninteracting
atoms obeying the Fermi-Dirac statistics. Substituting $\Psi_1=\Psi e^{iEt/\hbar}$ leads 
to the following eigenequation for the one particle solution $\Psi$, 
\begin{equation}
\frac{\hbar^2}{2m}[-\nabla^2+(\frac{\ell\alpha_0}{R^2})^2 r^2+2i(\frac{%
\ell\alpha_0}{R^2}) \partial_\phi]\Psi =E\Psi  \label{sch1}
\end{equation}
where the external trap $V_1(r)$ is chosen such that $V_{eff}=0$. After
rescaling the radial coordinate,  $r= x R$, and using the ansatz $\Psi=\xi(r)e^{i{q}\phi}e^{ik_z z}$ we obtain the solution in the form of a confluent hypergeometric function 
\begin{eqnarray}
 \xi(x)&=&  x^{|{q}|} e^{-\frac{|\ell |\alpha_0}{2}x^2}  \nonumber \\ && \!\!\!\!\!\!\!\!\!\!\!\!\!\!\!\!\!\!
{_1F_1}[\frac{1+|{q}|}{2}-(\frac{\epsilon}{4|\ell |\alpha_0}+\frac{q}{2}),|{q}|+1;|\ell |\alpha_0 x^2]
\label{sol}
\end{eqnarray}
where $\epsilon=(E-\frac{\hbar^2 k_z^2}{2m})\frac{2mR^2}{\hbar^2} $. 
The result
presented in Eq.(\ref{sol}) differs from the standard Landau problem \cite{lanlifqm} 
for free electrons in a homogenous magnetic field in the sense
that in our case we have to take into account the boundary conditions at $r=R$ 
and the fact that with a particular choice of light beam we choose the
form of the vector potential whereas in the Landau case the vector potential
is not uniquely defined. It is interesting to note at this point the
asymmetry induced by the vector potential. This is clearly seen in 
Eqs. (\ref{sch1}) and (\ref{sol}) where the energy eigenvalue is shifted by $2\ell\alpha_0 {q}$. As
such, Eq. (\ref{sol}) is rather intractable. We can, however, obtain
analytical expressions for the eigenvalues in the limit $|\ell |\alpha_0\gg1$, where the energies are of the form 
\begin{equation}
\epsilon_{n,{q}}=2|\ell | \alpha_0 (2n+|{q}|+{q}+1)  \label{strong}
\end{equation}
with $n=0,1,2,...$ and ${q}=...-2,-1,0,1,2...$. This is indeed the Landau result, since for 
$|\ell | \alpha_0\gg1$ the boundary conditions play a less important role. 

With the discrete energy levels given by Eq.(\ref{strong}) we are now in a
position to calculate thermodynamic potentials, such as the free energy, as
a function of the effective magnetic field. The free energy is defined as 
\begin{eqnarray}
F=N\eta-\int\frac{Z(E)}{e^{(E-\eta)/kT}+1}dE \label{free_ex}
\end{eqnarray}
where $Z(E)$ is the number of states with energies less than $E$, $\eta$ is
the chemical potential, $N$ the number of atoms, $k$ the Boltzmann constant
and $T$ the temperature. In order to proceed we need an expression for the
function $Z(E)$ using the discrete energy levels in Eq. (\ref{strong}). In
addition we assume a continuous atomic spectrum in z-direction. The resulting
function is 
\begin{equation}
Z(E)=\frac{\sqrt{2m}}{\hbar}\frac{L_z}{2\pi} \sum_{n,{q}} [E-\epsilon_{n{q}}]^
\frac{1}{2}  \label{number}
\end{equation}
where the sum is over $n$ and ${q}$ such that $Z(E)$ is real and $L_z$ is the
length of the cloud in z-direction. In the limit of $\varepsilon_0=\frac{\eta}{
\hbar\Omega}\gg1$ and with the rescaled temperature $\theta=\frac{kT}{
\hbar\Omega}$ we calculate, after carefully examining the sum in Eq. (\ref
{number}) \cite{lanlifsp1}, the free energy which becomes
\begin{eqnarray}
F&=&N\eta-A\left(\frac{\varepsilon_0^\frac{3}{2}}{24}+\frac{\pi^2}{3072}\frac{
\theta^2}{\sqrt{\varepsilon_0}}+\frac{4}{35}\varepsilon_0^\frac{7}{2}+\frac{\pi^2}{
96}\varepsilon_0^\frac{3}{2}\theta^2 - \right. \nonumber \\  
&& \!\!\!\!\!\!\!\!\!\!\!\!\!\!\! \left. \theta\frac{3}{4\pi} \sum_{s=1}^\infty \frac{\sqrt{s}\cos[\pi\varepsilon_0 s-\frac{\pi}{4}]-
\frac{\lambda}{\sqrt{s}} \cos[\pi\varepsilon_0 s-\frac{3}{4}\pi]}{(-1)^s s^2 \sinh[2\pi^2\theta s]}\right)
\label{freen}
\end{eqnarray}
where $A=\frac{\sqrt{2m}}{\hbar}\frac{L_z}{12\pi}(\hbar\Omega)^\frac{3}{2}$,
$\lambda=\frac{2^{11/2}15}{19 \pi^{3/2}}$ and $\hbar\Omega=\hbar^2|\ell |\alpha_0/2mR^2$. 
Using the free energy it is straightforward to calculate thermodynamic properties such as the specific heat 
\begin{equation}
C=-T \frac{\partial^2F}{\partial T^2},  \label{C}
\end{equation}
see Fig. (2), which consequently has an oscillating term as well. From Eq.(\ref{freen}) we see that for $\theta \ll 1$ 
the specific heat is linear in $\theta$ whereas the oscillating behavior stems from the 
$1/\hbar\Omega$ dependence in the cos-term.

\begin{figure}[tbp]
\center{
\includegraphics[width=6.5cm]{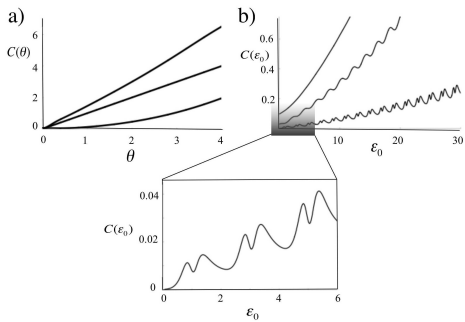}
\caption{a) The specific heat as a function of the rescaled temperature $\theta$. The lowest line corresponds to $\varepsilon_0=1$, middle one to $\varepsilon_0=10$ and the top one to $\varepsilon_0=30$. b) The specific heat as a function of the inverse effective magnetic field, $\varepsilon_0=\eta/\hbar\Omega$ where the three curves correspond to $\theta=0.1, 0.5$ and $1.0$ (top one). The inset shows a magnification of the region for small $\varepsilon_0$ and for $\theta=0.1$.}}
\end{figure}
In order to have a dominating oscillating part $\theta$ must not be too
large, $\theta\lesssim 1$, otherwise the oscillating term is exponentially
damped. With present experimental cooling and trapping techniques
temperatures of the order of $\theta/\varepsilon_0=T/T_F\sim 0.1$ are readily
achievable. Hence an $\varepsilon_0=\eta/\hbar\Omega$ of the order of one would
be preferable. For a homogeneous cloud in a cylindrical trap $\varepsilon_0$ is
given by $\varepsilon_0=(N\frac{R}{L_z} 3\pi)^\frac{2}{3}/(\alpha_0 \ell)$
which can become small if one considers a large aspect ratio trap and taking
into account the fact that the term $\alpha_0\ell$ can reach large values of
the order of $100$. In Fig. (2) we show the specific heat calculated from
Eq. (\ref{C}) where $C$ is in units of $k\frac{\sqrt{2m\hbar\Omega}}{\hbar}
\frac{L_z}{2\pi}$. In calculating  $C$ we used the exact equation (\ref{free_ex}) for $F$
to show the full range of the $\varepsilon_0$-dependence.

In this Letter we have shown how light with an orbital angular momentum can be
used to create an effective magnetic field in a degenerate gas of electrically 
neutral atomic fermions. As an example on how the slow light can be used we calculated the
free energy for the trapped degenerate fermions and found that the atomic gas shows
a de Haas-van Alphen type behavior where oscillations in thermodynamic
properties depend on the inverse effective magnetic field strength. There are other intriguing
phenomena such as the quantum Hall effect which can be studied 
using cold fermionic gases and slow light with an angular 
momentum. 
In addition, if the collisional interaction between the atoms is taken into account slow light 
can be used to study the magnetic properties of a superfluid atomic Fermi gas \cite{regal04}.
Recent advances in spatial light modulator technology enables us to consider rather
exotic lightbeams \cite{mcgloin03}. This will allow us to study the effect of different forms
of vector potentials in quantum gases. In particular the combined dynamical
system of light and matter could give important insight into gauge theories
in general. It is certainly tempting to push the analogy further and
study phenomena from high energy physics in ultracold samples of atoms.

\begin{acknowledgements}
This work was supported by the Royal Society of Edinburgh, the Royal Society of London 
and the Lithuanian State Science and Studies Foundation. 
Helpful discussions with E. Andersson, M. Babiker, S. Barnett, J. Courtial, M. Fleischhauer, 
U. Leonhardt, M. Lewenstein, M. Ma\v{s}alas, J. Ruseckas and L. Santos are gratefully acknowledged.
\end{acknowledgements}

\end{document}